\documentclass[twocolumn,showpacs,preprintnumbers,amsmath,amssymb,floatfix]{revtex4}
\usepackage{graphicx}
\usepackage{dcolumn}
\usepackage{bm}

\newcommand{\be}{\begin{equation}}
\newcommand{\ee}{\end{equation}}
\newcommand{\bea}{\begin{eqnarray}}
\newcommand{\eea}{\end{eqnarray}}
\newcommand{\barr}{\begin{eqnarray}}
\newcommand{\earr}{\end{eqnarray}}
\newcommand{\rar}{\rightarrow}
\newcommand{\pdup}{p_\uparrow}

\newcommand{\pimp}{\pi^- + \pdup \rar \pi^0 + X}

\newcommand{\xf}{x_{\mathrm F}}

\newcommand{\pdupp}{\pdup + p \rar \pi^0 + X}
\begin{document}

\title{
Indication on the universal hadron substructure ---
constituent quarks.
}

\author{V. V. Mochalov}%
\affiliation{Institute for High Energy Physics,
 Protvino, Moscow Region, 142281, Russia}

\author{S. M. Troshin}%
 \email{troshin@mx.ihep.su}
\affiliation{Institute for High Energy Physics,
 Protvino, Moscow Region, 142281, Russia}

\author{A. N. Vasiliev}%
\affiliation{Institute for High Energy Physics,
 Protvino, Moscow Region, 142281, Russia}


\begin{abstract}
The universality of single-spin asymmetry on inclusive $\pi$-meson
production is discussed. This universality can be related to
the hadron substructure --- constituent quarks.
\end{abstract}
\pacs{13.88.+e, 12.90+b}
\maketitle
Polarization experiments give us an unique opportunity to probe the
nucleon internal structure. While spin averaged cross-sections can be
calculated within acceptable accuracy, current theory of strong
interactions can not describe large spin asymmetries and polarization.
Polarization is a precision tool for
measuring the electroweak parameters, spin dependent structure
functions etc. After establishing the fact that the nucleon spin
is not described by simple summing of the quark spins, the
study of gluonic and orbital momentum contribution
to it is very important and intriguing.

\begin{figure}[hbt]
\centering
\resizebox{7cm}{!}{\includegraphics*{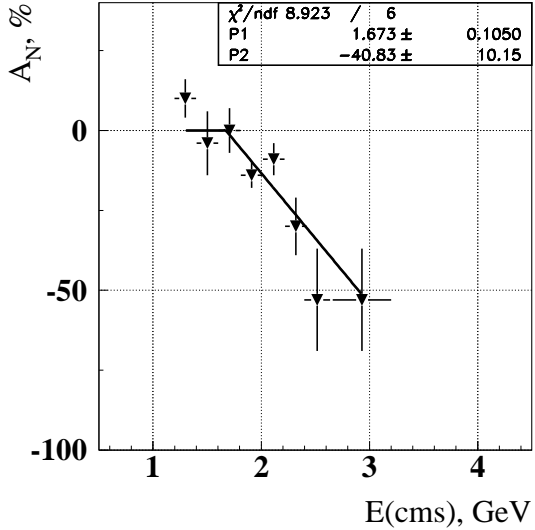}}
\caption{
The dependence of the single-spin asymmetry $A_N$ on   $\pi^0$-meson center
of mass energy in the reaction $\pimp$ in the central region at 40~GeV
(\cite{protv88}).
}
\label{fig:proza_cms1}
\end{figure}

\begin{figure}[hbt]
\centering
\resizebox{8cm}{!}{\includegraphics*{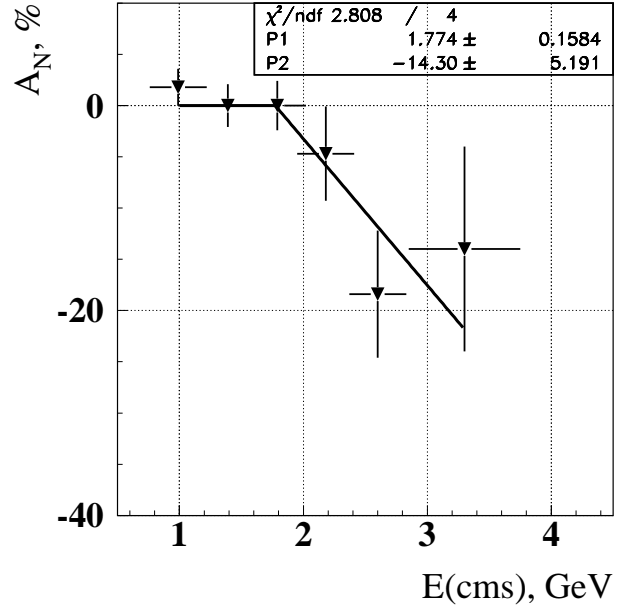}}
\caption{
The dependence of the single-spin asymmetry $A_N$ on   $\pi^0$-meson center
of mass energy in the reaction $\pimp$  in the polarized target
fragmentation region at 40~GeV
(\cite{proza40}).
}
\label{fig:proza_cms2}
\end{figure}

 Unexpected large values of  single spin asymmetry (SSA)
in inclusive $\pi$-meson production are real challenge
to current theory because perturbative Quantum Chromodynamics
predicts small asymmetries decreasing with transverse momentum.
Various models were developed to explain
results from E704 (FNAL), PROZA-M and FODS (both Protvino) and
several BNL experiments.
Most of the models analyse experimental data in terms of
 $\xf$ and/or $p_T$.
To investigate the dependence of SSA on a secondary meson production
angle, the measurements  in the reaction $\pimp$
were carried out at the PROZA-M experiment (Protvino) at 40~GeV pion
beam in the two different kinematic regions:
at Feinman scaling variable $\xf \approx 0$ \cite{protv88} and
in the polarized target fragmentation region \cite{proza40}.
The papers \cite{proza40,proza_scaling} reported that the asymmetry
of inclusive $\pi^0$
production in the reaction $\pimp$  begins to grow up at the same
centre of mass energy $E_0^{cms} \approx 1.7$~GeV.
The result is presented in { Figs.~\ref{fig:proza_cms1}, \ref{fig:proza_cms2}}.
Nevertheless from this statement we can not make the
conclusion whether the SSA behaviour
depends on a beam energy or not.

\begin{figure}[htb]
\centering
\resizebox{8cm}{!}{\includegraphics*{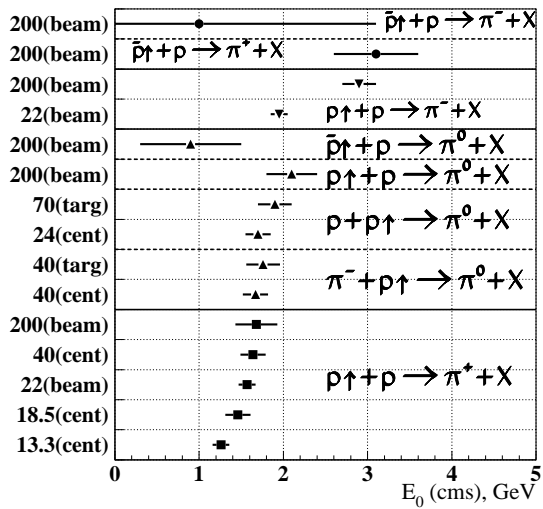}}

\caption {Center of mass energy values where  the pion asymmetry
begins to grow up for different experiments.
The energy along the Y-axis is in GeV; $cent$ -- corresponds to experiments in the central
region ( $x_f\approx0$), $targ$ -- the polarized target
fragmentation region; $beam$ -- the polarized beam fragmentation region.}
\label{fig:summary}
\end{figure}

The $\pi^+$ asymmetry in  E704 experiment
(200~GeV proton beam) \cite{e704fragm} and in E925 experiment
(BNL, 22~GeV)\cite{e925}   begins to rise up at different values of $\xf$
($\xf^0 \approx 0.18$ for E704 and $\xf^0 \approx 0.46$ for E925).
It was also found that the asymmetry for these two experiments
begins to grow up at the same longitudinal or full energy in the centre
of mass system, $E_0^{cms} \approx 1.6$~GeV. It happened to be
surprisingly the same energy as for the PROZA-M experiment.
The comprehensive analysis of all fixed target polarized experiments
of inclusive $\pi$-meson production was done in  \cite{proza_scaling}.
The result of the analysis is presented in  {Fig.~\ref{fig:summary}}.

The main conclusion is that the asymmetry begins
to grow up at the same center of mass energy $E_0^{cms}=1.5$ to $2.0$~GeV for
most of the experiments in the energy range between 13 and 200~GeV.
The analysis was done only for those experimental data where a transverse
momentum $p_T$ was greater than 0.5~GeV/c to exclude very soft interactions.
We did not include the experiments when the asymmetry was close to zero.
The conclusion is valid for all $\pi^+$ and $\pi^0$ asymmetries.
We have to mention that $\pi^-$ production seems to contradict to this.
We can explain this fact that $\pi^-$-meson at
small $x_F$ can be produced not only from the valence $d$-quark but also from
other channels. The interference of different channels is also responsible for
asymmetry cancellation in $\pi^0$ and $\pi^-$ production in the
central region. In the reaction $\pimp$ in the central region we found
significant asymmetry in the contrary to the $\pdupp$ reaction. If in
the $\pdupp$ reaction the asymmetry is
cancelled because of different channel interference
from a polarized and non-polarized proton, in the $\pi^-\pdup$ collisions
the valence $u$-quark from a polarized proton combining with the valence
$\bar{u}$-quark from $\pi^-$ gives the main contribution to $\pi^0$
production, while other channels are suppressed.)

In  this scheme the asymmetry behaviour in ${\bar{p}_{\uparrow}p}$
interactions in $\pi^+$ and $\pi^-$ production should be inversed in
comparison with the $\pdupp$ data. The result from E704
experiment \cite{e704_anti} is consistent with this model.
The asymmetry of $\pi^+$-production begins to grow up at the same value
$E_0^{cms} \approx 2.9$~GeV as for $\pi^-$ in reaction $\pdupp$, and
the asymmetry in the reaction ${\bar{p}_{\uparrow}+p \rar \pi^- +X}$ begins
to grow up at small value $E^0_{cms}$.

We may conclude that the meson asymmetry produced by valence
quark begins to grow up at the same universal energy $E^0_{cms}$.

The obtained universality of the value $E^0_{cms}$ can manifest
the presence of the universal substructures  in the hadrons --- constituent
quarks. The concept of constituent quark \cite{gmz,mor} has
been  used extensively since the very beginning of the quark era but
has  just obtained recently a  possible
direct experimental evidence at Jefferson Lab \cite{petron}.

A particular model for single spin asymmetries which used the
constituent quark concept in the hadron interaction
was proposed in \cite{spcon}.
The constituent quark appears
as a quasiparticle, i.e. as current valence quark surrounded by
the cloud of quark-antiquark pairs of different flavors, i.e. they are
structured hadron-like objects (cf. \cite{mor}).
SSA in the model is due to an
 orbital angular momentum of quarks
inside the constituent quark:
spin of constituent quark, e.g. $U$-quark
 is given
by the  sum:
\[
J_U=1/2=S_{u_v}+S_{\{\bar q q\}}+L_{\{\bar q q\}}=
1/2+S_{\{\bar q q\}}+L_{\{\bar q q\}}.
\]
On the grounds of the experimental data for polarized DIS
the conclusion was made  that the significant part of the spin
of constituent quark in the model should be associated with
 the orbital angular momentum
of the current quarks inside the constituent one \cite{spcon}.
In the model SSA reflects internal structure
of the constituent quarks and is proportional to the orbital angular
momentum of current quarks inside the constituent quark.
Evidently, SSA related to the internal orbital momentum
will be non-zero only  when the constituent quark will be excited
and broken up.
The  value $E^0_{cms}$ can be related then to the
minimal energy which is needed for constituent quark excitation and its
dissolution. In this approach it is natural that this energy is universal
since it is adherent to the properties of the constituent quarks. It should be related
anyway to the scale of chiral symmetry breaking $\Lambda_\chi^2$.

Thus the revealed scaling dependence of asymmetry can be interpreted as
 another indication of the presence of constituent quarks in the hadrons.
\section*{Acknowledgements}
We are grateful to N. E.~Tyurin for the interest to the work and fruitful discussions.
This work was partially supported by Russian Foundation for
Basic Research grant 03-02-16919.

\end{document}